\documentclass{ws-procs10x7-mod}
\usepackage{balance}
\usepackage[english]{babel}
\newcolumntype{d}[1]{D{.}{.}{#1}}

\usepackage{bm}
\usepackage{euscript,amsmath}

\newcommand{\lsim}{
\mathrel{\hbox{\rlap{\hbox{\lower4pt\hbox{$\sim$}}}\hbox{$<$}}}}

\makeindex
\begin{document}
\begin{flushright}
{\large DESY 06-159\\
September 2006}
\end{flushright} 
\vspace{2.0truecm}
\begin{center}
\boldmath
\large\bf A BRIEF ACCOUNT OF $B \to K^\ast \ell^+ \ell^-$ DECAY
 IN SOFT-COLLINEAR EFFECTIVE THEORY
 \unboldmath
\end{center}

\vspace{0.9truecm}
\begin{center}
\bf Ahmed Ali\\[0.1cm]
{\sl Theory Group, Deutsches Elektronen-Synchrotron
 DESY,\\ Notkestrasse 85, 22603 Hamburg, Germany}
\end{center}

\vspace{8truecm}

\begin{center}
{\bf \sl Presented at the XXXIII International Conference on High Energy
 Physics,\\ Moscow, Russian Federation, 26th July -- 2nd August 2006.\\
To appear in the Proceedings (Publishers: World Scientific, Singapore)}
\end{center}

\vfill

\newpage
\thispagestyle{empty}
\title{A BRIEF ACCOUNT OF \boldmath{$B \to K^\ast \ell^+ \ell^-$} DECAY
 IN SOFT-COLLINEAR EFFECTIVE THEORY}

\author{A. ALI$^*$}

\address{Theory Group, Deutsches Elektronen-Synchrotron
 DESY,\\ Notkestrasse 85, 22603 Hamburg, Germany\\
$^*$E-mail: ahmed.ali@desy.de}


\twocolumn[\maketitle\abstract{A brief account of the study of rare B decay
 $B \to K^\ast
\ell^+ \ell^-$ using soft-collinear effective theory (SCET)
 is presented. Theoretical underpinning  of this work is
a factorization formula, derived to leading power in $1/m_b$ and
 valid to all orders in $\alpha_s$. 
Partially integrated branching ratio in the dilepton squared mass range
$1~\mbox{GeV}^2 \le q^2 \le 7~\mbox{GeV}^2$
and  the forward-backward (FB) asymmetry of the leptons are calculated.
For the zero-point of the FB asymmetry, we get
$q_0^2=(4.07^{+0.16}_{-0.13})
~\mbox{GeV}^2$. The scale-related uncertainty of $q_0^2$ is improved
compared to the earlier estimate of the same.
 }
]
\section{Factorization in SCET}
 The emergence of an effective theory,
called soft-collinear effective theory (SCET)~\cite{Neubert:2005mu}, provides a
systematic  way to deal with the perturbative strong interaction
effects in  B decays in the heavy-quark expansion.  SCET
has been used extensively in the so-called heavy-to-light
transitions in $B$ decays.
In particular, this framework was 
 used to prove the factorization of radiative
$B \to K^* \gamma$ decay at leading power in $1/m_b$ and to all
orders in $\alpha_s$~\cite{Neubert05,Kim03}.
In a recent paper~\cite{Ali:2006ew},
summarized below, the related decay $B \to K^\ast \ell^+ \ell^-$
has been studied using the SCET approach.

 For the $b \to s$ transitions, the weak effective Hamiltonian can be written
 as~\cite{Beneke01}
\begin{equation}\label{Heff}
H_{eff}=-\frac{G_F}{\sqrt{2}}V_{ts}^\ast V_{tb} \sum_{i=1}^{10} C_i
(\mu) Q_i (\mu)~,
\end{equation}
neglecting terms proportional to $V_{us}^\ast V_{ub}$
and  using the unitarity of the CKM matrix.

Restricting  to the kinematic region where the  $K^\ast$
 meson  can be viewed approximately as
a collinear particle, a
factorization formula for the decay amplitude of $B \to K^* \ell^+\ell^-$,
to leading power in $1/m_b$ and all orders of $\alpha_s$, has been derived
 in SCET~\cite{Ali:2006ew}:
\begin{eqnarray}
\langle K_a^\ast \ell^+ \ell^- \vert H_{eff} \vert B \rangle &=& T^I_a(q^2)
\zeta_a(q^2) + \sum_{\pm} \int_0^\infty \frac{d\omega}{\omega} \nonumber\\
&&\hspace*{-3cm}\times\phi^{B}_{\pm}(\omega) \int_0^1 du ~\phi_{K^\ast}^{ a}(u)T^{II}_{a,\pm}
(\omega, u,q^2)~,
\end{eqnarray}
where $a=\parallel,\perp$ denotes the polarization of the $K^\ast$
meson. The functions $T^I$ and $T^{II}$ are perturbatively
calculable; $\zeta_a(q^2)$ are the soft form factors defined in SCET while
$\phi^{B}_{\pm}(\omega)$ and $\phi_{K^\ast}^{a}(u)$ are the light-cone distribution
amplitudes (LCDAs) for the B and $K^\ast$ mesons, respectively. 
The expression (2) coincides {\it formally} with the one obtained 
 by Beneke et al.~\cite{Beneke01} in $O(\alpha_s)$ accuracy, using the QCD
 factorization
approach~\cite{BBNS}. We calculate the 
partial dilepton invariant mass spectrum and the  forward-backward (FB)
asymmetry, and compare our results with the existing
 data~\cite{BaBar,Belle}
 and the earlier
theoretical analysis~\cite{Beneke01}.
\section{$B \to K^\ast \ell^+ \ell^-$ in SCET}
As SCET contains two kinds of collinear fields, called hard-collinear
and collinear fields, normally an intermediate effective theory,
$\mbox{SCET}_I$, is introduced which contains only soft and
hard-collinear fields. While the final effective theory, called
$\mbox{SCET}_{II}$, contains only soft and collinear fields. One undertakes
a two-step matching from $QCD \to \mbox{SCET}_I \to
\mbox{SCET}_{II}$~\cite{Bauer:2002aj}.
\subsection{QCD to $\mbox{SCET}_I$ matching}
In $\mbox{SCET}_I$, the $K^\ast$ meson is taken as a hard-collinear
particle. The matching from QCD to $\mbox{SCET}_I$
at leading power is expressed as
\begin{eqnarray}
H_{eff} \to &&\hspace*{-0.1cm}-\frac{G_F}{\sqrt{2}} V_{ts}^\ast V_{tb} \left (
\sum_{i=1}^4 \int \! ds~\widetilde{C}_i^A (s) J_i^A (s) \right .\nonumber\\
&&\hspace*{-1.2cm} + \sum_{j=1}^4
\int \! ds \int \! dr~\widetilde{C}_j^B (s,r) J_j^B
(s,r) \nonumber \\
&&\hspace*{-1.2cm} \left . +\int \! ds \int \! dr \int \! dt
~\widetilde{C}^C(s,r,t) J^C(s,r,t)\right ),
\end{eqnarray}
where $\widetilde{C}_i^{(A,B)}$ and $\widetilde{C}^C $ are Wilson coefficients
 in the position space. 
The operators $J_i^A$ and $J_i^B$ represent the cases that the
lepton pair is emitted from the $b \to s$ transition currents, while
$J^C$ represents the diagrams in which the lepton pair is
emitted from the spectator quark of the B meson. Their explicit 
expressions are given in our paper~\cite{Ali:2006ew}. It is more convenient
 to define the Wilson
coefficients in the momentum space.
 The corresponding coefficient functions are
 called $C_i^A(E)$, $C_j^B(E,u)$, and $ C^C(E,u)$,
with $E\equiv n\cdot v {\bar n}\cdot P/2$ and the velocity of the B meson is
defined as $v=P_B/m_B$. To get the
order $\alpha_s$ corrections to the decay amplitude, we need 
 the Wilson coefficients $C_i^A$ to one-loop level and
$C_j^{B}$ and $C^{C}$ to tree level.
\subsection{$\mbox{SCET}_I \to \mbox{SCET}_{II}$ matching
and SCET matrix elements}
 One may define the matrix
elements of the A-type $\mbox{SCET}_I$ currents as non-perturbative
input since the non-factorizable parts of the form factors are all
contained in such matrix elements~\cite{Bauer:2002aj,Neubert04B}.
Thus~\cite{Neubert05}:
%
\begin{equation}
\begin{aligned}
\hspace*{-1.5cm}\langle K^\ast \ell^+ \ell^- \vert J_1^A \vert B \rangle\\
&\hspace*{-2.0cm}=-2E\zeta_\perp(g^{\mu\nu}_\perp - i\epsilon^{\mu\nu}_\perp )
\varepsilon^{*}_{\perp\nu} \bar{\ell}\gamma_\mu \ell~,\\
&\hspace*{-2.5cm}\langle K^\ast \ell^+ \ell^- \vert J_2^A \vert B
\rangle\\
&\hspace*{-2.0cm} =-2E \zeta_\parallel
\frac{n^\mu}{n\cdot v} \bar{\ell}\gamma_\mu \ell~,
\end{aligned}
\end{equation}
where $g^{\mu\nu}_\perp \equiv g^{\mu\nu}-(n^\mu {\bar n}^\nu+{\bar
n}^\mu n^\nu)/2$ and $\epsilon^{\mu\nu}_\perp\equiv
\epsilon^{\mu\nu\rho\sigma}v_\rho n_\sigma/(n\cdot v)$, and
 we use the convention $\epsilon^{0123}=+1$. The matrix elements
of the other two $A$-type currents, 
$ \langle K^\ast \ell^+ \ell^- \vert J_3^A \vert B \rangle $ and
$\langle K^\ast \ell^+ \ell^- \vert J_4^A \vert B \rangle$, are
obtained from the above matrix elements by the replacement
$ \bar{\ell}\gamma_\mu \ell \to \bar{\ell}\gamma_\mu \gamma_5 \ell $,
respectively.

The B-type $\mbox{SCET}_I$ operators are matched onto the
$\mbox{SCET}_{II}$ operators $O_i^B$ ($i=1,...,4$). Their
matrix elements involve the meson
LCDAs, and two different $K^*$-distribution amplitudes
($\phi_{K^*}^\parallel(u,\mu)  $
for $\Gamma=1$ and $\phi_{K^*}^\perp(u,\mu)  $
for $\Gamma=\gamma_\perp$) with their corresponding decay constants
$f_{K^*}^\parallel$ and 
$f_{K^*}^\perp(\mu)$, respectively, are required.  
With the above LCDAs, one has~\cite{Ali:2006ew}
\begin{equation}
\begin{aligned}
\langle K^\ast \ell^+ \ell^- \vert C_1^B O_1^B \vert B \rangle\\
&\hspace*{-3.0cm}= -\frac{F(\mu)m_B^{3/2}}{4}(1-\hat{s})(g^{\mu\nu}_\perp -
i\epsilon^{\mu\nu}_\perp )\\
&\hspace*{-3.0cm}\times \varepsilon^{*}_{\perp\nu}\bar{\ell}\gamma_\mu \ell~ \phi_+^B
\otimes f_{K^\ast_\perp}\phi_{K^\ast_\perp} \otimes {\cal J}_\perp
\otimes C_1^B~,
\end{aligned}
\end{equation}
\begin{equation}
\begin{aligned}
\langle K^\ast \ell^+ \ell^- \vert C_2^B O_2^B \vert B
\rangle&=-\frac{F(\mu)m_B^{3/2}}{4}(1-\hat{s})\\
&\hspace*{-3.0cm}\times \frac{n^\mu}{n\cdot
v}\bar{\ell}\gamma_\mu \ell \phi_+^B \otimes
f_{K^\ast_\parallel}\phi_{K^\ast_\parallel} \otimes {\cal
J}_\parallel \otimes C_2^B~,
\end{aligned}
\end{equation}
where $\otimes$ represent convolution in the appropriate variables,
 $F(\mu)$ is related to the B meson decay constant $f_B$ up to higher
orders in $1/m_b$, and the jet functions
${\cal J}_i$ arise from the $\mbox{SCET}_I \to \mbox{SCET}_{II}$
matching, with ${\cal J}_1= {\cal J}_3\equiv{\cal J}_\perp$ and
${\cal J}_2= {\cal J}_4\equiv{\cal J}_\parallel$.
The matrix element of $C_3^B O_3^B$($C_4^B O_4^B$) can
be obtained by replacing the lepton current
$\bar{\ell}\gamma_\mu \ell$ on the right hand side of the above
equations by $\bar{\ell}\gamma_\mu \gamma_5 \ell$ and also
replacing $C_1^B
\to C_3^B$ ($C_2^B \to C_4^B$).

Finally, the C-type $\mbox{SCET}_I$ current is matched
onto the $\mbox{SCET}_{II}$ operator $O_C$, with its
Wilson coefficient defined  in the
momentum space, $D^C(\omega, u, \hat{s}, \mu) $, and
 we  define an auxiliary function
$D^C \equiv \widehat{D}^C/(\omega-q^2/m_b-i\epsilon)$. With this,
the matrix element of $O^C$ is obtained in SCET, with the
 result~\cite{Ali:2006ew}
\begin{eqnarray}
\langle K^\ast \ell^+ \ell^- \vert D^C O^C \vert B
\rangle&=&-\frac{F(\mu)m_B^{3/2}}{4}(1-\hat{s})\nonumber\\
&&\hspace*{-3.5cm}\times \frac{\bar{n}^\mu}{\bar{n}\cdot
v}\bar{\ell} \gamma_\mu \ell
 \frac{\omega \phi_-^B}{\omega-q^2/m_b-i\epsilon}
\otimes f_{K^\ast_\parallel} \phi_{K^\ast_\parallel} \otimes
\widehat{D}^C~.\nonumber\\
\end{eqnarray}
Since $\phi_-^B(\omega)$ does not vanish as $\omega$ approaches
zero, the integral $\int \!
d\omega~\phi_-^B(\omega)/(\omega-q^2/m_b)$ would be divergent if
$q^2\to 0$. This endpoint singularity will violate the
$\mbox{SCET}_{II}$ factorization, and we should restrict the kinematic
 region so that the invariant mass of the
lepton pair is not too small, say $q^2\geq 1~\mbox{GeV}^2$.
\subsection{Resummation of logarithms in SCET}
The two-step matching procedure $\mbox{QCD} \to
\mbox{SCET}_I \to \mbox{SCET}_{II}$ 
introduces two matching scales, $\mu_h \sim m_b$ at which QCD is
matched onto SCET$_I$, and $\mu_l \sim \sqrt{m_b \Lambda_h}$ at which
SCET$_I$ is matched onto SCET$_{II}$ ($\Lambda_h$
represents a typical hadronic scale).  The large
logarithms due to different scales are resummed 
using the renormalization-group equations (RGE) of SCET$_I$ to
evolve from $\mu_h$ to  $\mu_l$.

For the A-type SCET currents, only the
scale $\mu_h$ is involved.
 For the B-type currents, the RGE of SCET$_I$ can
be obtained by calculating the anomalous dimensions of the relevant
SCET operators~\cite{Neubert04}, and the
matching coefficients at any scale $\mu$ can be obtained by an evolution from
the matching scale $\mu_h$.
 The resulting evolution equation has been
solved numerically~\cite{Ali:2006ew}.

Finally, for the C-type SCET current $J^C$, its anomalous dimension
just equals the sum of the anomalous dimensions of the $K^\ast$
meson LCDA $\phi_{K^*}$ and the B meson LCDA $\phi_-^B$. As the
evolution equation of $\phi_-^B$ is still unknown, 
 the  perturbative logarithms for the $J^C$ current are not
resummed. 
 Numerically the contribution from the $J^C$ current to the decay
amplitude in $B \to K^\ast \ell^+ \ell^-$ is small. Furthermore, 
the $J^C$ current is  irrelevant for the FB asymmetry of the
charged leptons. 
\section{Dilepton invariant mass and FB asymmetry}
The dilepton invariant mass spectrum and the
FB asymmetry in $B \to K^* \ell^+\ell^-$
have the following expressions in SCET:
\begin{equation}
\begin{aligned}
\frac{d Br}{d q^2} &= \tau_B \frac{G_F^2 \vert
V_{ts}^\ast V_{tb} \vert^2}{96\pi^3} \left
(\frac{\alpha_{em}}{4\pi} \right )^2 m_B^3 \vert \lambda_{K^\ast}\vert\\
&\hspace*{-0.8cm}\times (1-\frac{q^2}{m_B^2})^2 {\cal N}(q^2,\zeta_\perp^2,
\zeta_\parallel^2)~,\nonumber
\end{aligned}
\end{equation}
%
%
\begin{equation}\label{DAFB}
\begin{aligned}
\frac{d A_{FB}}{d q^2}&=
\frac{-6 (q^2/m_B^2) \zeta_\perp^2 Re({\cal C}_{9}^\perp)
{\cal C}_{10}^\perp }{ {\cal N}(q^2,\zeta_\perp^2,\zeta_\parallel^2)}~.
\end{aligned}
\end{equation}
where the  function
 ${\cal N}(q^2,\zeta_\perp^2,\zeta_\parallel^2)$ is defined as
\begin{eqnarray}
{\cal N}(q^2,\zeta_\perp^2,\zeta_\parallel^2)&\equiv& 4\frac{q^2}{m_B^2}
 \nonumber\\
& & \hspace*{-2.5cm}\times\zeta_\perp^2  (\vert
{\cal C}_{9}^\perp \vert^2 + ({\cal C}_{10}^\perp)^2 )
+\zeta_\parallel^2(\vert {\cal C}_{9}^\parallel \vert^2 + ({\cal
C}_{10}^\parallel)^2 )~.
\end{eqnarray}
 The expressions for the "effective" Wilson coefficients
 ${\cal C}_9^{\perp,\parallel}$ and ${\cal
C}_{10}^{\perp,\parallel}$ in SCET,
 valid at leading power in $1/m_b$ and to
all orders in $\alpha_s$,  can be seen
in our paper~\cite{Ali:2006ew}. 
%
\subsection{Numerical results}
%
  We use the radiative $B \to K^\ast
\gamma$ decay rate,
 which has been measured quite precisely, to normalize the
soft form factor at $q^2=0$, obtaining
 $\zeta_\perp(0)=0.32 \pm 0.02$.
 The longitudinal soft form factor
$\zeta_\parallel(q^2)$ is obtained from the full QCD form factor
$A_0^{B\to K^\ast}(q^2)$, estimated using the LCSRs~\cite{Ball05},
yielding $\zeta_\parallel(0)=0.40 \pm 0.05$. For both the soft form
factors, we assume that their $q^2$-dependence 
 can be reliably obtained from the LCSRs~\cite{Ball05}.
The rest of the input parameters and the values for the Wilson coefficients
can be seen in our paper~\cite{Ali:2006ew}.
  We obtain
\begin{eqnarray}
\int \limits_{1\mbox{\scriptsize ~GeV}^2}^{7\mbox{\scriptsize ~GeV}^2} d q^2
\frac{d Br(B^+ \to K^{\ast +} \ell^+ \ell^-)}{dq^2}&=&\nonumber\\
&&\hspace*{-6.0cm}(2.92^{+0.57}_{-0.50}
\vert_{\zeta_\parallel}~^{+0.30}_{-0.28} \vert_{\mbox{\scriptsize CKM}}
~^{+0.18}_{-0.20})\times 10^{-7}~,
\end{eqnarray}
making explicit the uncertainties from the soft form factor
$\zeta_\parallel$ and the CKM factor $\vert V_{ts}^* V_{tb} \vert$.
The last error reflects the uncertainty
due to the variation of the other input parameters and the residual scale
dependence. 
For $B^0$ decay, the branching ratio is about $7\%$ lower due to the
lifetime difference (ignoring the small isospin-violating corrections from
the matrix elements).
One of the
Belle observations \cite{Belle} of our interest is
\begin{eqnarray}
\int \limits_{4\mbox{\scriptsize ~GeV}^2}^{8\mbox{\scriptsize ~GeV}^2} d q^2
\frac{d Br(B \to K^\ast \ell^+ \ell^-)}{dq^2}&=&\nonumber\\
&&\hspace*{-5.5cm}(4.8^{+1.4}_{-1.2}
\vert_{\mbox{\scriptsize stat.}}\pm 0.3 \vert_{\mbox{\scriptsize syst.}}
\pm 0.3 \vert_{\mbox{\scriptsize model}})\times 10^{-7},
\end{eqnarray}
for which we predict~\cite{Ali:2006ew}
 $(1.94^{+0.44}_{-0.40}) \times 10^{-7}$, which is smaller
than the published Belle data by a factor of about 2.5. However,
 BaBar collaboration measures the total branching ratio of
$B \to K^\ast \ell^+ \ell^-$ to be \cite{BaBar}
$(7.8^{+1.9}_{-1.7}\pm 1.2) \times 10^{-7}$,
which is about a factor 2 smaller than the Belle measurement of the
 same \cite{Belle},
$(16.5^{+2.3}_{-2.2}\pm 0.9 \pm 0.4) \times 10^{-7}$. Clearly, more
data is required to test the theory precisely.
%
\begin{figure}[t]
\begin{center}
\unitlength 0.8mm
\begin{picture}(80,50)
\put(0,0){\includegraphics[width=0.47\textwidth]{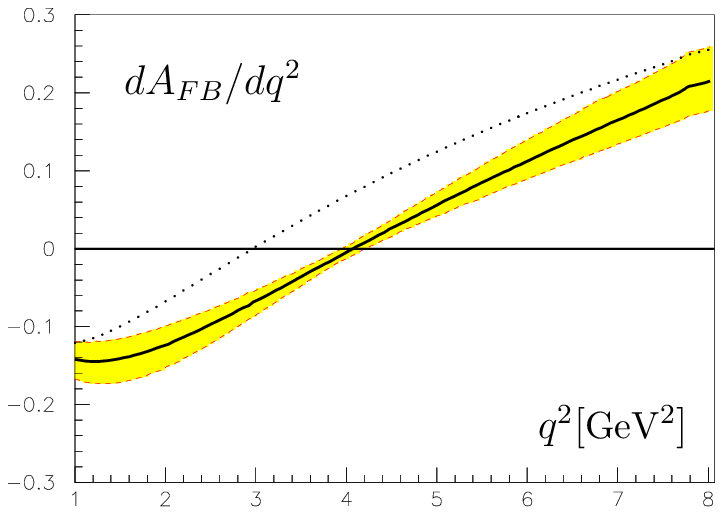}}
\end{picture}
 \caption{The differential FB
 asymmetry $d A_{FB}(B \to K^\ast \ell^+ \ell^-)/d q^2$. 
Solid line corresponds to
 the input parameters taken at their central values, while the gray
band reflects the uncertainties from input parameters
and scale dependence. The dotted line represents the LO predictions
 (from Ref.~4 .)}
\end{center}
\end{figure}

 The zero of the FB asymmetry is determined by
$Re({\cal C}_{9}^\perp)=0$.
Including the order $\alpha_s$ corrections, our analysis estimates the
zero-point of the FB asymmetry to be~\cite{Ali:2006ew}
\begin{equation}
q^2_0=(4.07^{+0.16}_{-0.13})~ \mbox{GeV}^2~,
\end{equation}
of which the scale-related uncertainty is
 $\Delta(q_0^2)_{\rm scale}=^{+0.08}_{-0.05}$
 GeV$^2$ for the range
 $m_b/2 \leq \mu_h \leq 2 m_b$ together with the jet function scale
$\mu_l=\sqrt{\mu_h \times 0.5~\mbox{GeV}}$. This is to be compared with
the result given in Eq.~(74) of Beneke et al.~\cite{Beneke01}, also obtained in the
absence of $1/m_b$ corrections:
$ q^2_0=(4.39^{+0.38}_{-0.35})~ \mbox{GeV}^2$. Of this the largest single
uncertainty (about $\pm 0.25~ \mbox{GeV}^2 $) is attributed to the scale
dependence.  The difference in the
estimates of the scale dependence of $q_0^2$ here and by Beneke
 et al.~\cite{Beneke01} is mainly due to the incorporation of the SCET
 logarithmic resummation~\cite{Ali:2006ew}
 and to a lesser extent to the different
 (scheme-dependent)
definitions of the effective form factors for the SCET currents.
Power corrections in $1/m_b$  are probably
comparable to the $O(\alpha_s)$ corrections~\cite{Beneke01};
it remains to be seen how a model-independent calculation of the same
effect the numerical value of $q_0^2$.  

\end{document}